\documentclass[10pt]{sigplanconf}

\usepackage{amsmath}
\usepackage[english]{babel}
\usepackage[utf8]{inputenc}
\usepackage[T1]{fontenc}
\usepackage{type1ec}
\usepackage{graphicx}
\usepackage{subfigurequickfix}
\usepackage{setspace}
\usepackage{url}

\usepackage{hyperref}
\usepackage{listings}
\usepackage{parcolumns}
\usepackage{color}
\usepackage{amsfonts}
\usepackage{colortbl}
\usepackage{booktabs}
\usepackage{amsmath}
\usepackage{algorithmicx}
\usepackage{algorithm}
\usepackage{algpseudocode}
\usepackage{tabularx}
\usepackage{textcomp}
\usepackage{soul}
\usepackage{footmisc}
\usepackage{changepage}
\usepackage[dvipsnames]{xcolor}
 \usepackage[all]{nowidow}

\newcommand{\aspas}[1]{{``#1''}}

\newcommand{\mcode}[1]{\texttt{#1}}

\definecolor{jsbackground}{RGB}{253,246,227}
\definecolor{jscomment}{RGB}{147,161,161}
\definecolor{jsstring}{RGB}{42,161,152}
\definecolor{jskeywords}{RGB}{133,153,0}
\definecolor{jsidentifier}{RGB}{88,110,117}
\definecolor{jsndkeywords}{RGB}{38,139,210}

\lstdefinelanguage{JavaScript}{
    keywords = {typeof, new, true, false, catch, return, null, catch, switch, if, in, while, do, else, case, break, +, <, >, *, /, =, ===, ==, +=, -=, *=, /=, <=, >=, !=, ++, --},
    ndkeywords = {function, var, this, prototype, Array, String, Number},
    sensitive = true,
    alsoletter = {<>+-=*/},
    alsodigit = {-},
    comment = [l]{//},
    morecomment = [s]{/*}{*/},
    morestring = [b]',
    morestring = [b]"
}

\lstdefinelanguage{HTML5}[]{HTML}{
    sensitive=false,
    alsoletter={-},
    morekeywords={section, ng-app, ng-controller, ng-submit, header, ng-model, placeholder, autofocus},
    tag=[s]
}

\lstdefinestyle{jsstyle}{
    backgroundcolor=\color{jsbackground},
    commentstyle=\itshape\color{jscomment},
    stringstyle=\color{jsstring},
    keywordstyle=\color{jskeywords},
    ndkeywordstyle=\color{jsndkeywords}\bfseries,
    numberstyle=\tiny\color{jsstring},
    identifierstyle=\color{jsidentifier},
    basicstyle={\scriptsize\ttfamily},
    xleftmargin={0.75cm},
    numbers=left,
    stepnumber=1,
    firstnumber=1,
    numberfirstline=true,
    breakatwhitespace=false,
    breaklines=true,
    captionpos=b,
    keepspaces=true,
    numbers=left,
    numbersep=5pt,
    showspaces=false,
    showstringspaces=false,
    showtabs=false,
    tabsize=2,
    upquote=true
}

\newcommand{\angularjs}{{\sc AngularJS}}
\newcommand{\AngularJS}{{\sc AngularJS}}

\begin{document}

\setlength{\pdfpageheight}{\paperheight}
\setlength{\pdfpagewidth}{\paperwidth}

\conferenceinfo{PLATEAU '16,}{November 01 2016, Amsterdam, Netherlands}
\copyrightyear{2016}
\copyrightdata{978-1-4503-4638-2/16/11}
\copyrightdoi{3001878.3001881}

%\CopyrightY ear{2016}
%\setcopyright{acmcopyright}
%\conferenceinfo{PLA TEAU '16,}{November 01 2016, Amsterdam, Netherlands} \isbn{978-1-4503-4638-2/16/11}\acmPrice{\$15.00} \doi{http://dx.doi.org/10.1145/3001878.3001881}

% Uncomment the publication rights you want to use.
%\publicationrights{transferred}
%\publicationrights{licensed}     % this is the default
%\publicationrights{author-pays}

%\titlebanner{AngularJS in the Wild: Results of a Survey with 460 Developers}        % These are ignored unless
%\preprintfooter{\tiny 7th Workshop on Evaluation and Usability of Programming Languages and Tools (PLATEAU)}   % 'preprint' option specified.

\title{AngularJS in the Wild: A Survey with 460 Developers}

\authorinfo{Miguel Ramos\and \mbox{Marco Tulio Valente}}
           {UFMG, Brazil}
           {\{miguel,mtov\}@dcc.ufmg.br}
\authorinfo{Ricardo Terra}
           {UFLA, Brazil}
           {terra@dcc.ufla.br}
\authorinfo{Gustavo Santos}
           {RMoD Team, INRIA, France}
           {gustavo.santos@inria.fr}
\maketitle

\begin{abstract}
To implement modern web applications, a new family of JavaScript frameworks has emerged, using the MVC pattern. Among these frameworks, the most popular one is \AngularJS, which is supported by Google. In spite of its popularity, there is not a clear knowledge on how \AngularJS\ design and features affect the development experience of Web applications. Therefore, this paper reports the results of a survey about \AngularJS, including answers from 460 developers. Our contributions include the identification of the most appreciated features of \AngularJS\ (e.g., custom interface components, dependency injection, and two-way data binding) and the most problematic aspects of the framework (e.g., performance  and implementation of directives).
%\textcolor{red}{escrever mais}
%, common practices to deal with performance problems, and general and technical causes of these problems.
\end{abstract}


% general terms are not compulsory anymore,
% you may leave them out
%\terms
%term1, term2


\section{Introduction}
\label{chap:introduction}

JavaScript is a fundamental piece of modern Web applications.
It was initially designed as a scripting language to extend web pages with small executable code. However, the language is used nowadays to construct a variety of complex systems~\cite{kienle2010s,saner2015}. As a result, we are observing the birth of new technologies and tools---including JavaScript libraries and frameworks---to solve common problems faced in the development of such applications.
For example,  frameworks following the Model-View-Controller (MVC) architecture pattern (or variations of it) are widely used nowadays, including systems such as \angularjs, {\sc Backbone.js}, and {\sc Ember.js}. Among these frameworks, \angularjs\ is probably the most popular one. This fact is evidenced by comparing the number of Google searches (the most queried framework since 2013),
the number of contributors in GitHub %(1,418 contributors vs.~555 and 283 in Ember and Backbone, respectively), %(~\ref{img:frameworkscontributions}),
and the increasing number of questions and answers in Stack Overflow (the framework with more Q\&A since mid-2013).

However, despite the increasing practical interest on \AngularJS, \ul{there is no clear knowledge on how the design and features proposed by this framework affect the development experience of JavaScript software}. More specifically, it is not clear what are the most appreciated features of \AngularJS, what are the main problems faced by developers when using the framework, and which aspects of \AngularJS\ can be improved. Answers to these questions are important to different developers. First, 
developers {\em who use} \AngularJS\ can learn how to improve this usage and also how to avoid bad \AngularJS\ programming practices. Second, developers {\em who do not} use JavaScript MVC frameworks can understand the benefits and problems related to these frameworks, by reviewing the case of \AngularJS.  Third, MVC framework builders can use our results to design more powerful and usable frameworks.

This paper reports the results of a survey with 460 developers, when we collected their perceptions about \AngularJS. We reveal the relevant features of the framework, e.g.,
custom components, dependency injection, and two-way data binding.
We also shed light on the most frequent problems faced by \AngularJS\ developers, e.g.,
due to the complexity of the API to declare directives.

The remainder of the paper is organized as follows. Section~\ref{chap:background}
introduces \AngularJS. Section~\ref{chap:angularsurvey} documents the survey design and 
Section~\ref{sec:angularresults} presents the survey results.
Threats to validity are presented in Section~\ref{sec:threadsinangular}.
Section~\ref{sec:relatedwork} discusses related work and
Section~\ref{chap:conclusion} concludes.

\section{AngularJS in a Nutshell}
\label{chap:background}

\noindent In this section, we briefly describe the key components of \AngularJS. A basic understanding of these components is important to interpret our survey results.\\[-2ex]

\noindent{\em Modules:} An \AngularJS\ application is a set of modules, which act as containers to the different parts of the application. Modules can also depend on other modules. \\[-2ex]

\noindent{\em Services:} \AngularJS\ services are objects that encapsulate code related with a specific concern. They are instantiated only once by factories or constructor functions. The created singleton object is shared by the components that depend on it (e.g., controllers, directives, filters, and other services). Typically, \AngularJS\ services are stateless objects with a set of methods that deal with specific concerns, such as server requests, manipulation of arrays, asynchronous operations, etc. \AngularJS\ also provides built-in services to deal with common concerns in Web applications, such as \mcode{\$http}, \mcode{\$filter}, and \mcode{\$timeout}.\\[-2ex]

\noindent{\em Templates:} In Web applications, HTML documents are parsed to generate the DOM (Document Object Model), which is the data structure that models the final document presented to users. \AngularJS\ supports DOM-based templates, which are written in HTML and contain proprietary elements and attributes to render the dynamic interface of  web applications. Listing~\ref{lst:templateexample} shows a sample template. It includes the definition of the document (\mcode{html} element in line~1)  and the markup for the main input of the application, which is represented by the \mcode{form} element (line~5).\\[-3.7ex]

\noindent
\begin{figure}[!t]
\centering
\begin{minipage}{\linewidth}
\begin{lstlisting}[
	language=HTML5,
	label={lst:templateexample},
	caption={Template sample}
]
<html lang="en" ng-app="todomvc">
...
	<header id="header">
		<h1>todos</h1>
		<form ng-submit="addTodo()">
			<input placeholder="What needs to be done?" ng-model="newTodo"/>
		</form>
	</header>
...
</html>
\end{lstlisting}
%\vspace{5pt}
\end{minipage}
\end{figure}

\noindent{\em Directives:} Directives are specific HTML markers used in templates to define the UI behavior. In Listing~\ref{lst:templateexample}, the \mcode{ng-app} attribute (line~1) is an \AngularJS\ directive that specifies the root node of the application. When directives are executed by \AngularJS\ they can modify the DOM structure and also register event handlers. During the compilation process, \AngularJS\ traverses the DOM  and searches for all directives. The directives are executed in order of priority generating the final DOM presented to users. Directives can (1) use the same scope of the parent element; (2) create a scope that inherits from the scope of the parent element; or (3) create a completely new scope. It is also possible to create custom directives. \\[-2ex]
%For example, it is possible to create HTML tags that are replaced with a more complex structure or custom attributes to modify the DOM or add event handlers in a specific way.\\[-1.5ex]

\noindent{\em Expressions:} \AngularJS\ expressions are delimited by double curly brackets (\mcode{\{\{expression\}\}}) or are the values of some directive attributes. Literals (e.g., arrays (\mcode{[]}), objects (\mcode{\{\}})), operators, and variables are examples of elements that can be used in expressions.
% \textcolor{red}{It is also possible to call functions and use the ternary operator (\mcode{?}).}\textcolor{green}{(It has to be maintained because gives the reason for which templates can become hard to read)}
% using the ? operator to say that it's hard to read is not convincing
Expressions are evaluated using a context represented by an object, called scope. Variables and functions used in expressions must be defined in the scope object. During the template compilation, when the \mcode{ng-app} directive is parsed, \AngularJS\ creates an object representing the main application scope, which is referenced as the \mcode{\$rootScope}. Since directives can define different scopes from \mcode{\$rootScope}, 
%However, new scopes that inherit from \mcode{\$rootScope} can be created by other directives. For this reason,
expressions from different parts of the template may be evaluated under different scopes. When the value of an expression changes, \AngularJS\ updates the view accordingly.\\[-2ex]

\noindent{\em Controllers:} \AngularJS\ controllers are used to initialize the state of an application and provide an interface to update it. Controllers are used with the \mcode{ngController} directive. When this directive is used in a template, it receives the name of a controller and the scope created by the directive is passed as a parameter to the specified controller. The controller must populate the scope object with properties and methods that are used when evaluating expressions.\\[-2ex]

\noindent{\em Digest Cycle:} \AngularJS\ constantly maintains in sync the state of the application with the view presented to the final user.
%\textcolor{green}{\st{As an example, suppose a simple AngularJS application providing a To-Do list. This application interface has a button to remove a task from the list. When it is clicked, the} \mcode{\st{removeTodo}} \st{method is executed and the task is removed from the internal array representing the tasks.}} \textcolor{red}{why not using listing 1 example?}\textcolor{green}{(Listing 1 is not adequate to describe this concept)
%\st{Moreover, without having to write additional code, the DOM is automatically updated to reflect the internal change.}}
The framework provides this synchronization by
%\textcolor{green}{\st{constantly observing the variables in the scope referenced by the template expressions.}
comparing the current value of all variables in the scope referenced by the template expressions with their previous values. When a change is detected, the framework adequately updates the DOM, in a process known as \emph{digest cycle}.

\section{Survey Design}
\label{chap:angularsurvey}

First, we performed a \emph{Mapping Study} (Section~\ref{sub:angularmappingstudy}) to gather information about \AngularJS. We relied on the results of this first study to construct the survey. The survey participants were selected among Stack Overflow users (Section~\ref{sub:angularsurveydesign}).
 
\subsection{Mapping Study}
\label{sub:angularmappingstudy}

We use a mapping study to gather information about the use of \AngularJS.
A mapping study is more flexible than a systematic review and it is recommended
for studying emergent fields or technologies~\citep{wohlin2012experimentation}.
%For example, a mapping study allows a more flexible criteria to include or exclude sources of information.
%We focus on the design of the framework, its current state of development, programming practices, developers' perceptions about the framework and common problems with its adoption.
For example, information about \AngularJS\ is primarily found in blogs, forums, and Q\&A sites.
Therefore, using the Google search engine, we initially focused on finding documents reporting the benefits and disadvantages of \AngularJS. To this purpose, we used search queries such as \aspas{\emph{the best features of AngularJS}} or \aspas{\emph{the bad parts of AngularJS}}.
We also used two other strategies to reach more sources of information. First, we recursively accessed the references in the sites already reviewed (a practice called \aspas{snowballing}). Second, we performed additional search queries for frequently mentioned topics. As an example, we searched for \aspas{\emph{transclusion directives}}, due to the frequency of references to this topic as a complex \AngularJS\ concept.
During the revision of blogs, we only considered well-written posts, by authors with experience in software development.   A complete list  of the posts we consider is available in a companion website.\footnote{\label{url-mapping}\url{https://github.com/aserg-ufmg/angularjs-survey}} 
%For each blog post, this document lists the URL, the date in which it was accessed, the name and the author's professional background. 
Finally, these documents were analyzed and classified by identifying topic trends, which were used for the survey~design.

\begin{figure*}[!t]
\centering
%\vspace{0.5cm}
\includegraphics[width=0.7\linewidth]{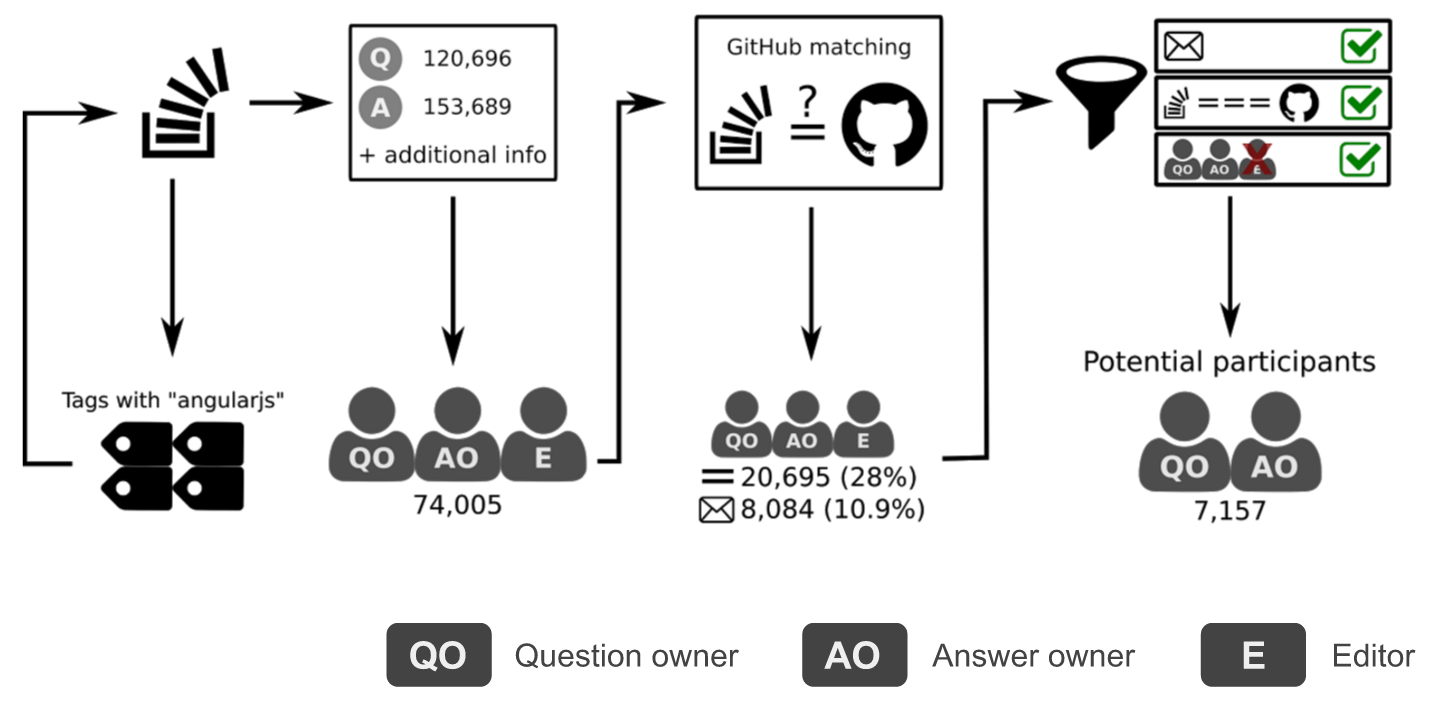}
%\vspace{-0.5cm}
\caption{Selecting the survey participants}
\label{fig:participants}
\end{figure*}

\subsection{Survey Construction and Participants}
\label{sub:angularsurveydesign}

As a result of the mapping study,
we built a 15-minute survey with 25 questions, organized in seven sections: (1)~background of the participants,
(2)~key characteristics of \AngularJS,
(3)~problems with \AngularJS\ templates,
(4)~debugging and testing,
(5)~development practices,
(6)~complex concepts and features, and
(7)~\AngularJS\ 2.0.
To avoid unreliable responses, we asked developers to skip the questions they did not feel confident to answer. All the scales used in the survey have an even number of points, to force participants to make a choice. In multiple choice questions, when the respondents could provide a response not included in the set of answers, we included an {\em Other} option.

To find participants, we used the Stack Exchange API\footnote{\url{https://api.stackexchange.com}} to search for \AngularJS\ developers in the Stack Overflow community. As illustrated in Figure~\ref{fig:participants}, we retrieved all existing tags with the substring \aspas{angularjs}. We retrieved 120,696 questions and 153,689 answers containing these tags. Next, we extracted data about three types of users: users who own a question (QO); users who own an answer (AO); or users who edited a question or answer (E). In this way, we collected a total of 74,005 users. To find their e-mail address, we matched each user nickname at StackOverflow with an equivalent nickname at GitHub. In this way, we obtained a total of 20,695 matched users (28\%). However, only 8,084 users (10.9\%) had public contact information at GitHub. In a last step, 7,157 users were marked as potential participants in the survey, because they have valid e-mail address and at least one operation with the collected answers that is not a simple edition.  These users were ranked considering the number of \AngularJS-related questions and answers at Stack Overflow. Each user received the following score: $S = \mathit{QO} + 3 * \mathit{AO}$, where $\mathit{QO}$ is number of questions owned by the user and $\mathit{AO}$ is the number of answers he owns. We gave an additional weight to answers since users who provide answers tend to have more experience than those who are looking for them.

We randomly selected 30 users from the middle of the rank, with scores between 9 and 24, to run a pilot survey.
Their feedback helped us to correct typographical errors and ambiguities in some questions. However, the most important change was in the structure of some questions. In the initial survey version, we used ranking questions, when survey respondents have to rank a list of items in order of importance. After this pilot study, we decided to change to rating questions (when respondents just have to rank each item in a scale ranging from 1 to 4) because some participants complained about the time to answer ranking questions. 
%Moreover, many participants interrupted the pilot survey in these questions. 
Due to these changes in the survey, the responses obtained during the pilot phase were discarded.

\begin{figure*}[!t]
\centering
\centering
\subfigure[JavaScript experience \label{fig:jsexperience}]{
\includegraphics[width=0.32\linewidth]{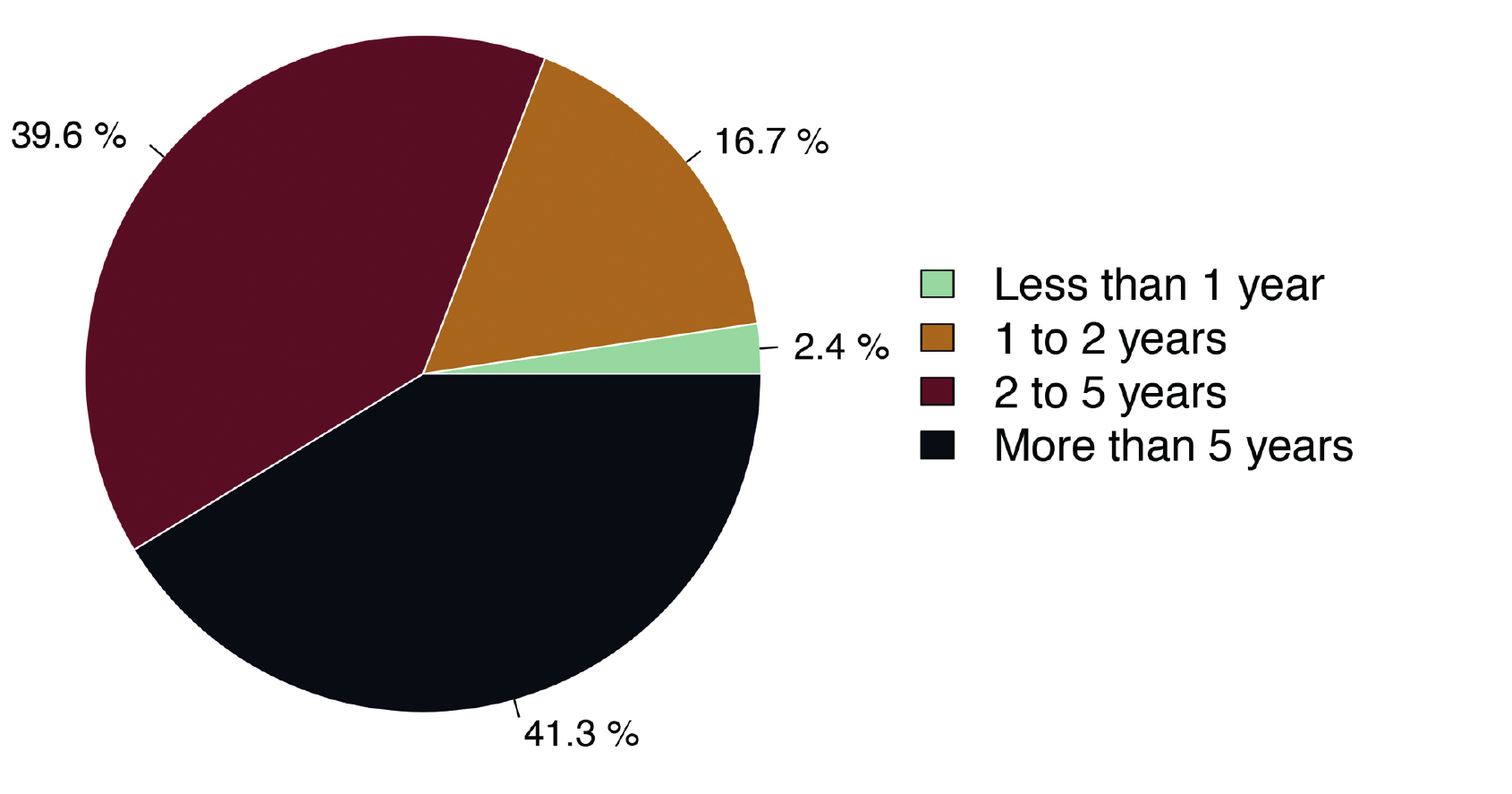}}
\subfigure[AngularJS experience \label{fig:angularexperience}]{
\includegraphics[width=0.32\linewidth]{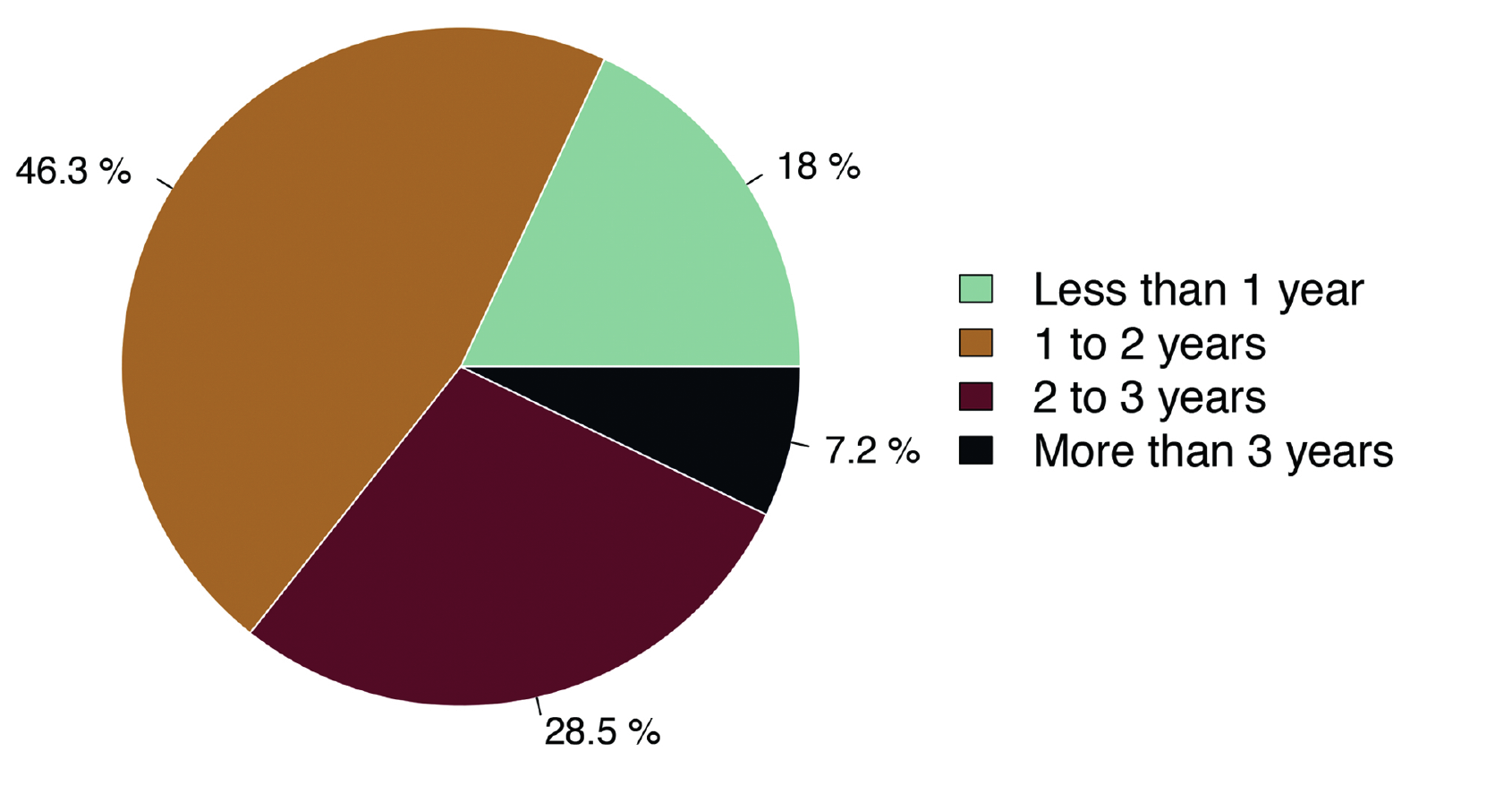}}
\subfigure[Largest application \label{fig:largestapplication}]{
\includegraphics[width=0.32\linewidth]{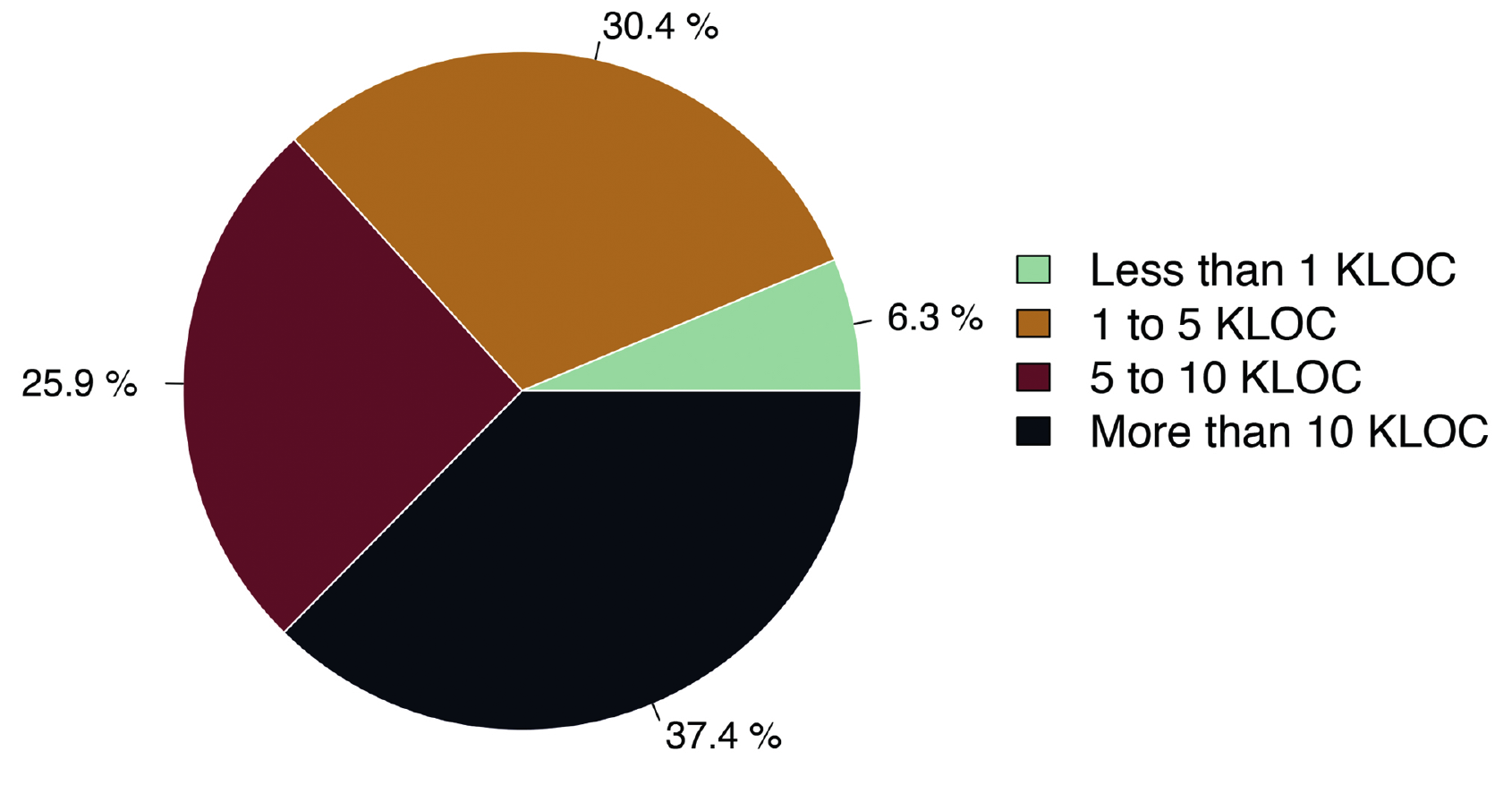}}
\caption{Respondents' background}
\label{fig:backgroundangular}
\end{figure*}

\begin{figure*}[!t]
\centering
%\vspace{0.5cm}
\includegraphics[width=0.67\linewidth]{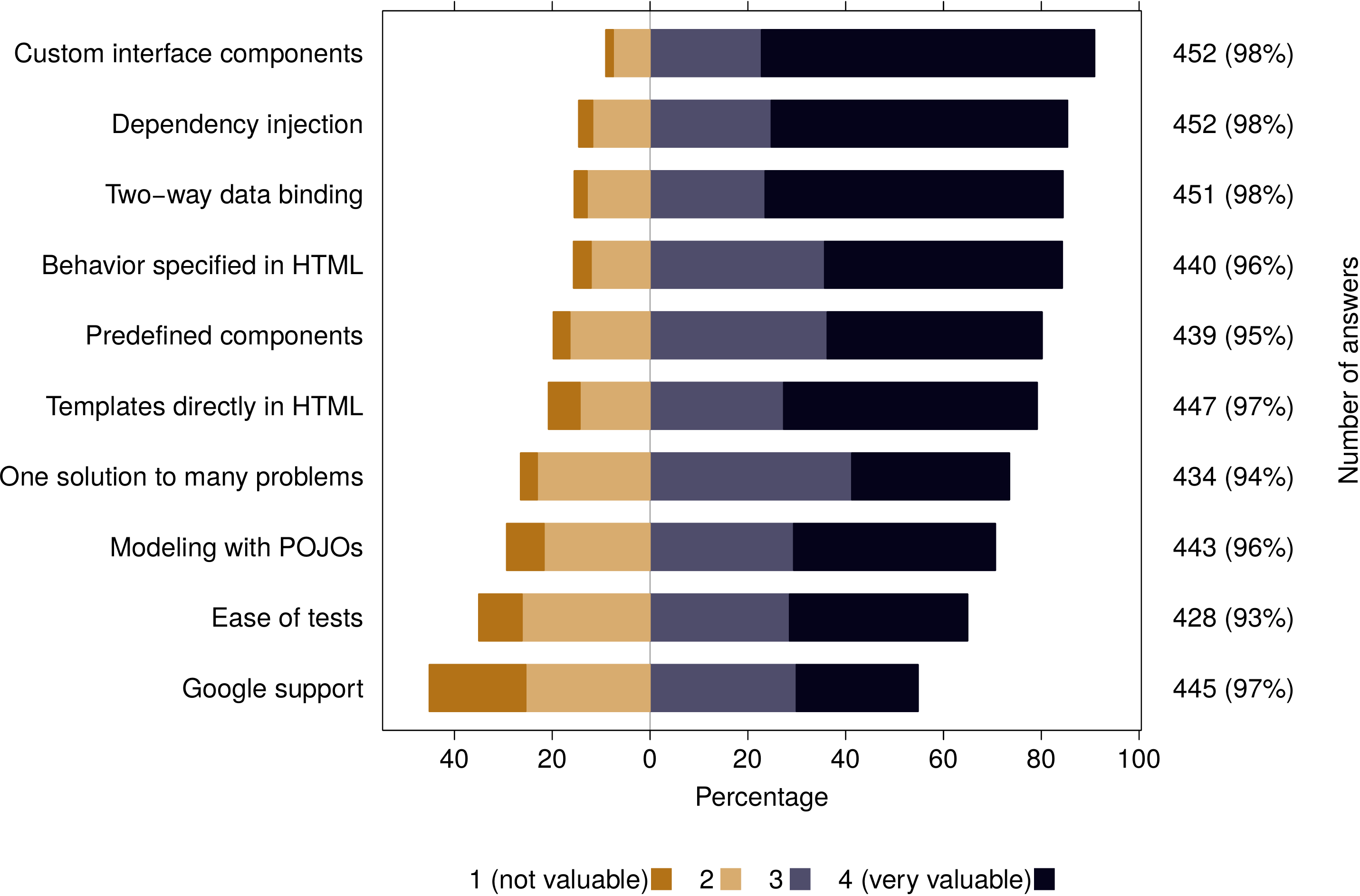}
%\vspace{-0.5cm}
\caption{Key features and characteristics of AngularJS}
\label{fig:keyfeatures}
\end{figure*}

The final version of the survey---which is also available in our companion website---was first sent to a group of 60 users. This time, we received complete responses and an improved response rate. Therefore, we extended the invitation to the rest of the users by daily emailing groups of nearly 700 users from the top of the ranking.
At one point, we decided to stop due to replies from developers saying that they had a limited experience with \AngularJS. In total, we sent the survey to 3,060 users with score between 3 and 831. We received 460 complete responses, representing a response rate of 15\%. The survey was open during approximately one month (from early November to early December 2015).

\section{Survey Results}
\label{sec:angularresults}

%In this section, we present the results of the survey. These results are examined in subsections that reflect the thematic structure of the questionnaire.

\subsection{Background}
\label{sub:angularresultsbackground}

Figure~\ref{fig:backgroundangular} reveals the participants background.
The majority of the respondents (97.6\%) have at least one year of experience in JavaScript (Figure~\ref{fig:jsexperience}),
and 74.8\% have one to three years of experience with \AngularJS\ (Figure~\ref{fig:angularexperience}).
For 37.4\% of the developers, the largest application implemented with \AngularJS\
has more than 10 KLOC (Figure~\ref{fig:largestapplication}).
Therefore, we can conclude that at least the participants are not novice \AngularJS\ developers.

\subsection{Key Characteristics and Features of AngularJS}
\label{sub:angularresultskeyfeatures}

We asked developers about the the following features:\\[-0.65cm]

\begin{enumerate}

\item \emph{Pre-defined components for code organization:} \AngularJS\ has different components to modularize code, which may help in separation of concerns.

\item \emph{Dependency injection:} This design pattern is used by \AngularJS\ to manage dependencies between components, to reduce coupling and increase testability.

\item \emph{Use of POJOs in model components:} In \AngularJS, models are implemented using Plain Old JavaScript Objects (POJOs). There is no need to extend proprietary classes, for example, to provide accessor methods.
%, which helps to directly deal with server data in JSON format.

\item \emph{Templates in HTML:} \AngularJS\ uses DOM-based templates to simplify data binding operations, event mapping, and updating of large interface components.

\item \emph{Support to custom components:} Custom directives can be used as a DSL to define reusable UI components.

\item \emph{Ease of writing tests:} \AngularJS\ provides the \mcode{ngMock} module to simulate logging operations, HTTP reqs, etc.

\item \emph{Two-way data binding:} \AngularJS\ provides synchronization between data in the view and in the model.

\item \emph{Use of HTML to declare UI behavior:} The UI, including its behavior, is defined in standard HTML documents.

\item \emph{One solution to manage many problems:} \AngularJS\ is an \aspas{opinionated framework}, meaning that its design handles common decisions related with Web apps.

\item \emph{Supported by Google:} This support may guarantee the evolution and constant maintenance of the project.
\end{enumerate}

\begin{figure*}[!t]
\centering
%\vspace{0.5cm}
\includegraphics[width=0.64\linewidth]{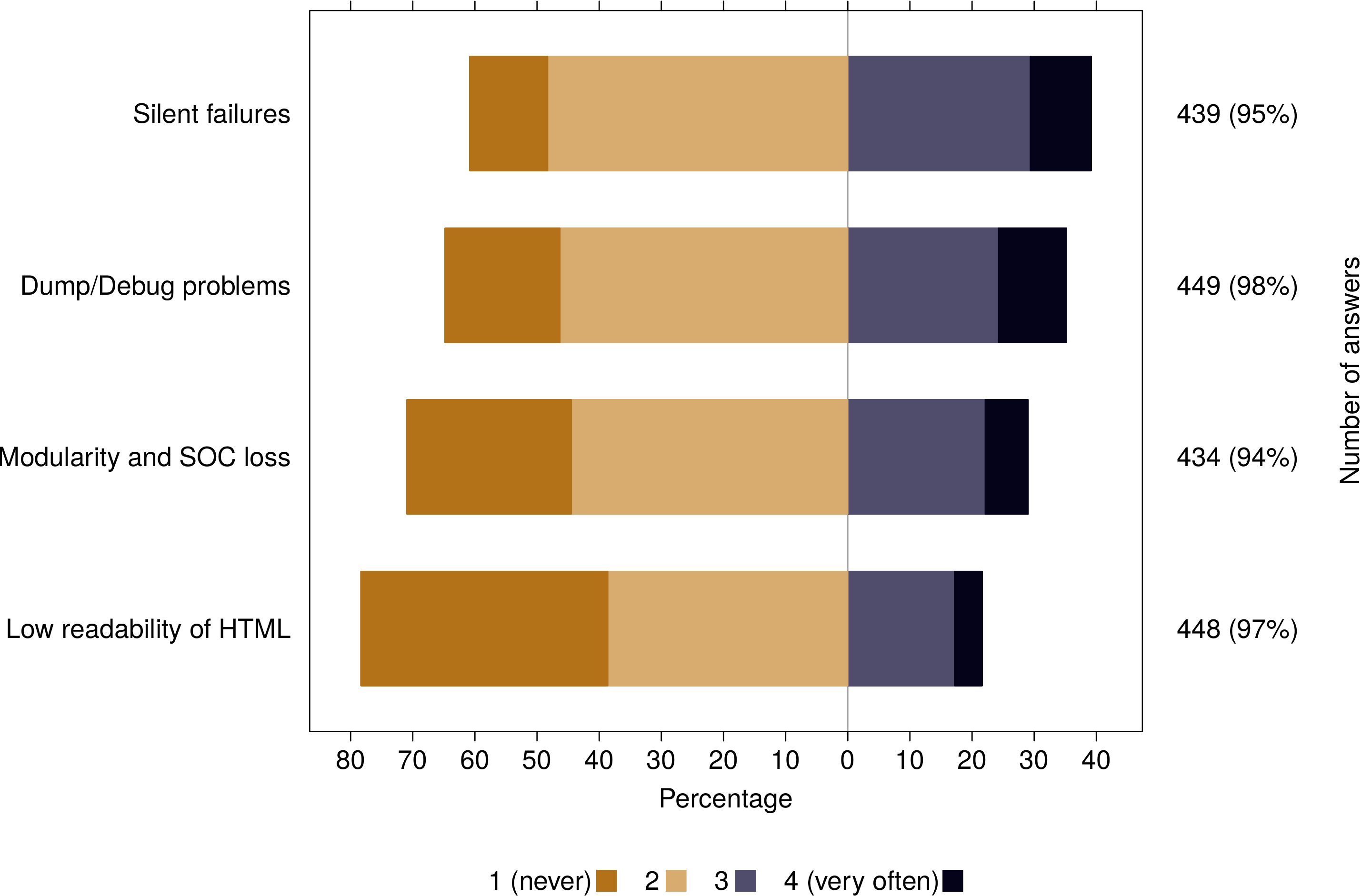}
%\vspace{-0.5cm}
\caption{How often these items represent a real problem caused by using code in \AngularJS\ templates?}
\label{fig:templateproblems}
\end{figure*}

Figure~\ref{fig:keyfeatures} reports the value given by developers to these characteristics and features, ranging from {\em not valuable} (score 1) to {\em very valuable} (score 4). At the right of the charts, we include the number of answers of each item. Each individual item was rated by at least 93\% of the participants. The top-3 features with more positive scores are the support to {\em custom components}, the  use of {\em dependency injection}, and the support for {\em two-way data binding}.
%These features are evaluated with scores 3 and 4 by, respectively, 90.9\%, 85.4\%, and 84.5\% of their respondents. By contrast, they
%have no value (score 1) to only 1.5\%, 2.9\%, and 2.7\% of the respondents.
The characteristic with the lowest number of positive answers is {\em Google support}, with 45.2\% of the respondents seeing it as having no value (score 1) or limited value (score 2). 
%The other two features with the lowest number of positive answers are \emph{ease of testing} and \emph{modeling with POJOs}.

\subsection{Code in HTML Templates}
\label{sub:codeintemplates}

In the mapping study, we detected four possible problems related to placing code in \AngularJS\ templates:

\begin{enumerate}

\item \emph{Silent failures:} In \AngularJS, when undefined functions or objects are used in templates no exceptions are raised. As a consequence, applications might fail silently.

\item \emph{Code hard to debug:} Since the code used in \AngularJS\ templates is not  pure JavaScript, it is not possible, for example, to define break points.

\item \emph{Low readability:} \AngularJS\ code might be spread all over the HTML document, hindering readability.

\item \emph{Modularity and Separation of Concerns:} The use of large amount of JavaScript code in HTML is often seen as bad smell~\citep{nguyen2012detection}, which might hinder separation of concerns.

\end{enumerate}

We asked developers whether these issues are real problems in their daily development. In this case,
the score 1 means the issue was never a problem and a score 4 means it is a very frequent problem.
As shown in Figure~\ref{fig:templateproblems},
none of the issues have a major detrimental impact, according to the respondents; they have at least 60\% of the answers in the low part of the scale (scores 1 or 2).

In a separate question, we asked if the developers had at least once used large amounts of logic in HTML templates and 26.3\% of them answered positively. Since this is not a recommended practice, we asked them to indicate possible reasons for their decision
%, including (a) lack of experience in JavaScript, (b) lack of experience in Web applications, (c) lack of experience in AngularJS and (d) AngularJS design, which might foster this practice. Figure~\ref{fig:largetemplatelogiccauses}\textcolor{green}{this figure is not in the document} shows the results.
The two most voted reasons are \AngularJS\ design (54.8\%) and the lack of experience in \AngularJS\ (43.5\%).
Lack of experience in Web architecture and in JavaScript were also voted (20.9\% and 4.3\%, respectively).
We also gave the respondents the possibility to indicate other reasons, which were provided by 31 developers. These reasons include, for example, tight deadlines, special cases, easiness or laziness, prototyping purposes, etc.

%\begin{figure}[htb]
%\centering
%\vspace{0.5cm}
%\includegraphics[width=0.7\linewidth]{img/angular/LargeTemplateLogicCauses.pdf}
%\vspace{-0.5cm}
%\caption{Reasons to place large amounts of code in HTML templates}
%\label{fig:largetemplatelogiccauses}
%\end{figure}

\subsection{Testing}
\label{sub:angulartesting}

\begin{figure*}[!t]
\centering
%\vspace{0.5cm}
\includegraphics[width=0.725\linewidth]{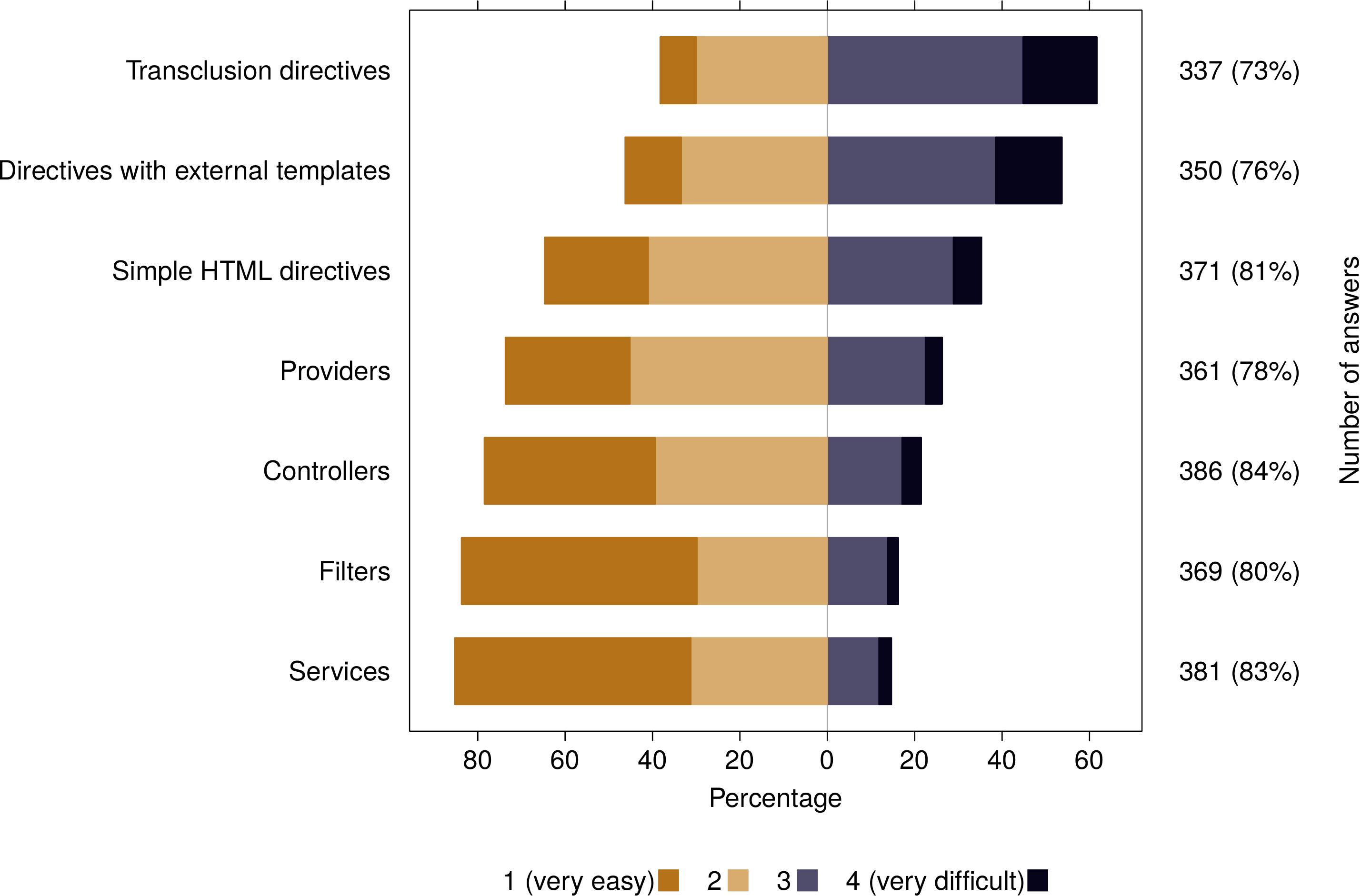}
%\vspace{-0.5cm}
\caption{How difficult is it to test these \AngularJS\ components?}
\label{fig:abstractionstestingcplxty}
\end{figure*}

First, we asked the participants to rate the frequency they make use of mocking (provided by the \mcode{ngMock} module) when testing their systems. From 441 answers,
72.8\% indicated they never or rarely use this module (scores 1 or 2). Possible reasons for this result include limited usefulness of \mcode{ngMock} features, unfamiliarity with the module, and few developers putting testing into practice. We also asked the participants to rank how complex is testing \AngularJS\ components, from very easy to very difficult.
As presented in Figure~\ref{fig:abstractionstestingcplxty},
{\em services}, {\em filters}, {\em controllers}, and {\em providers}
received the higher number of answers with scores 1 and 2.
The reason is that these components are very common and usually
do not include complex code or code that deals with complex APIs.
For example, most code in filters only make string transformations.
By contrast, the two components more difficult to test are {\em transclusion directives} and {\em directives with external templates}. Probably, developers find these directives more difficult to test because they demand a deeper understanding of \AngularJS\ concepts.
For example, when creating directives, there are different types of transclusion, different types of scopes, and different ways to interact with the DOM API.

\begin{figure*}[htb]
\centering
%\vspace{0.5cm}
\includegraphics[width=0.67\linewidth]{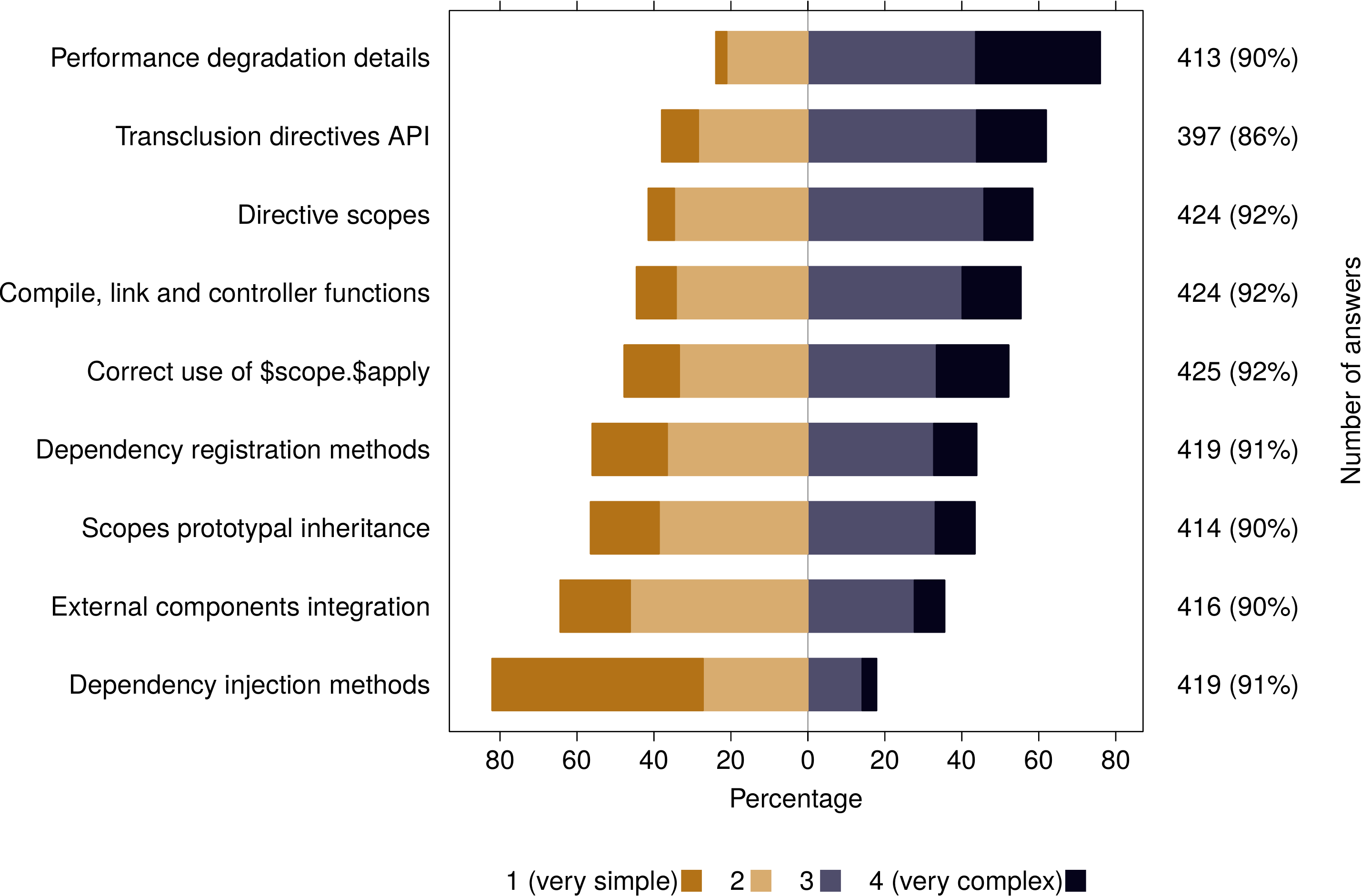}
%\vspace{-0.3cm}
\caption{How complex are the the following \AngularJS\ aspects and features?}
\label{fig:complexcharacteristics}
\end{figure*}

\begin{figure*}[htb]
\centering
%\vspace{0.5cm}
\includegraphics[width=0.75\linewidth]{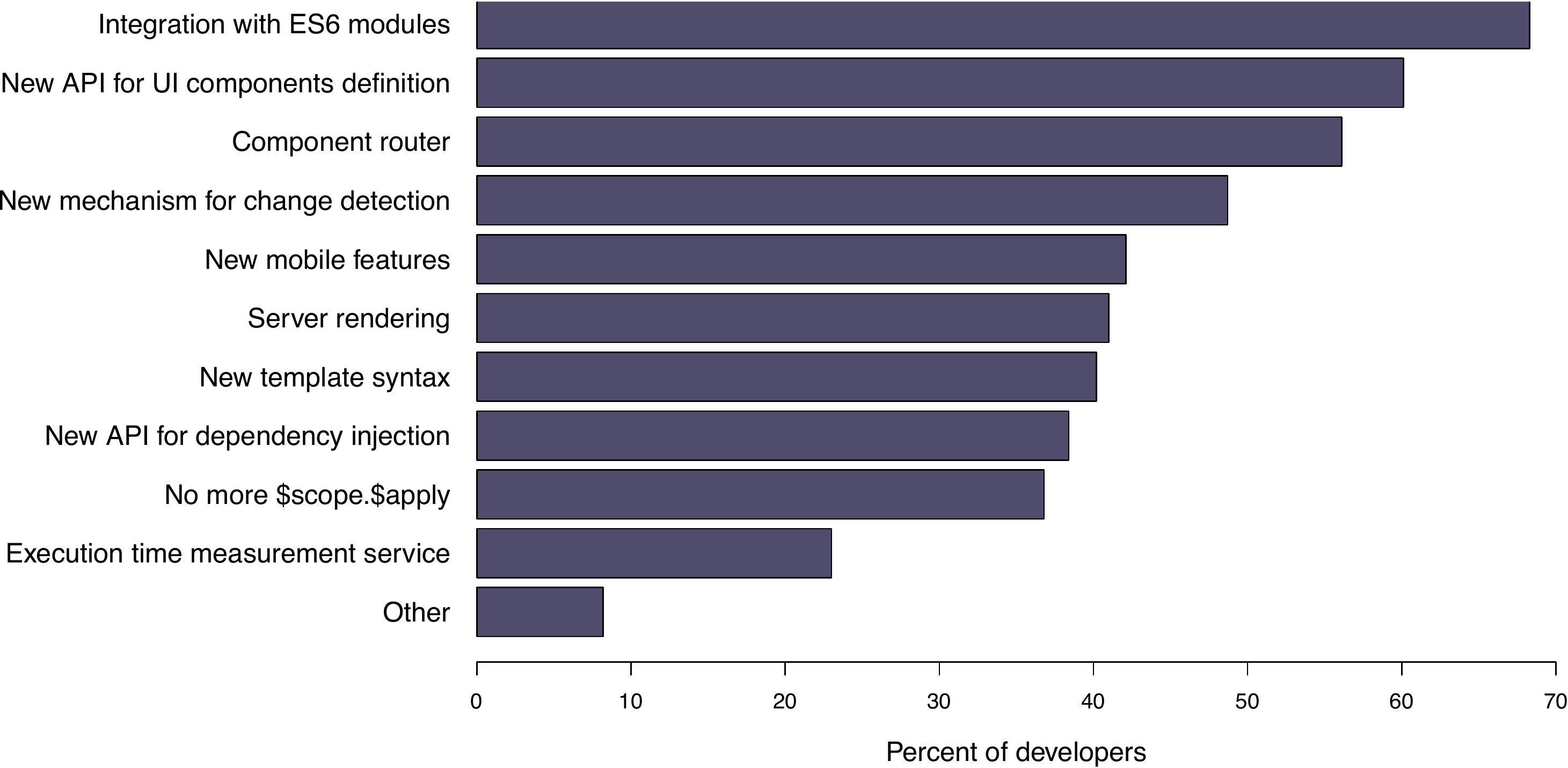}
%\vspace{-0.5cm}
\caption{Most expected \AngularJS\ 2.0 features}
\label{fig:mostexpectedangular2features}
\end{figure*}

\subsection{Complex Concepts and Features}
\label{sub:angularmastering}

We asked the participants to evaluate several characteristics of \AngularJS,
%, which were identified in the mapping study as dealing with or providing complex concepts.
which were originally identified in the mapping study as complex concepts.
The following characteristics were proposed:
%Using scores from 1 (very simple concept) to 4 (very complex concept), we asked the participants to rank the complexity of the following characteristics:
(1) use of different scopes in directives; 
(2) use of prototypes to simulate scope inheritance;
(3) different types of entities to register dependencies;
(4) compile, link, and controller functions (necessary to implement DOM-related directives);
(5) transclusion directives;
(6) correct use of \mcode{\$scope.\$apply} to manually trigger the digest cycle;
(7) tackling of performance degradation details;
(8) integration with external components and plug-ins; and
(9) correct usage of the syntax to inject dependencies (in order to avoid unexpected results when minifying the code).

Figure~\ref{fig:complexcharacteristics} summarizes the developers' classification
from very simple concept (score 1) to very complex one (score 4).
Tackling all the details that can lead to performance degradation was rated by 76\% of the participants as a complex task.
%, which confirms the importance of the survey about performance, presented in Section~\ref{chap:angularperf}.
The next three items in terms of complexity are transclusion directives,
the different scopes that can be used when building directives,
and the correct use of the functions \mcode{compile}, \mcode{link}, and \mcode{controller}.
All these items are somehow related with the implementation of directives.
The remaining items were rated as more simple than complex, mainly the
 integration with external components and  the use of the correct syntax to inject dependencies.
 
In a separate question, we asked the developers about the frequency they create directives.
As a result, 83.5\% of the participants answered they often or very often create their own directives. Despite this fact, many tasks and features related to the implementation of directives are usually considered complex by the survey participants.

\subsection{AngularJS 2.0}
\label{sub:angulartwo}

To reveal the most expected features or improvements in \AngularJS\ 2.0, we selected the following features that appeared in the mapping study: (1) the new API to define the main UI building blocks; (2) Zone.js (no more \mcode{\$scope.\$apply()}); (3) server rendering; (4) the new logging service, called diary.js; (5) new mobile features (e.g. support for touch event gestures); (6) the new template syntax; (7) the new change detection mechanism; (8) the new API for injecting dependencies; (9) integration with ECMAScript 6 (ES6) modules; (10) component router, which allows asynchronous loading.

Figure~\ref{fig:mostexpectedangular2features} indicates that the most expected feature is the integration with ES6 modules~(68.3\%). The second most expected feature is the new API to create UI components, which is expected because the current API is perceived by most developers as difficult to use and understand (see Section~\ref{sub:angularmastering}).
The third most expected feature is the component router, which allows the dynamic loading of UI components, followed by the new change detection mechanism.

\subsection{Key Findings and Implications}

The main findings of our survey are as follows: %\\[-0.85cm]

\begin{itemize}
\item Three characteristics of \AngularJS\ excel by the value that developers give to them: the ability to create UI components by means of custom directives, the use of dependency injection, and the ease to set up two-way data binding.
Therefore, JavaScript MVC framework builders can embrace these
characteristics and improve them by offering a more simple interface to build reusable UI
components (without exposing final users to internal concepts and decisions)
and by using  better mechanisms to detect changes in the model  (i.e., mechanisms that reduce the
number of details to be considered).

\item The two problems that arise more frequently regarding the use of code in \AngularJS\ templates are the emergence of silent failures and the difficulty to dump/debug the variables referenced in the HTML. New debug tools and techniques can then be developed to alleviate these problems faced by developers.

\item The two main reasons for placing large amounts of logic in templates are the lack of experience in \AngularJS\ and the design of the framework. This shows the importance of correctly training developers before they start to use \AngularJS\ on complex applications. It also reveals an opportunity for framework builders to investigate new framework designs.

\item The components that are more difficult to test in \AngularJS\ are directives, mainly the ones using transclusion; the remaining components (i.e., controllers, services, filters, etc.) are mostly considered easy to test. 
%\item although the creation of custom directives is one of the most appreciated characteristics of \AngularJS, the API that supports this feature is perceived as complex to use and understand by developers.
%\item Tackling all the details to avoid performance degradation is considered as one of the most complex characteristics of \AngularJS.
\end{itemize}

\section{Threats to Validity}
\label{sec:threadsinangular}

The first threat to validity relates to the execution of the mapping study. Due to the large amount of information on the Internet about \AngularJS, it is possible that literature addressing more specific topics about the framework or presenting different points of view was not included. 
 
Regarding the construction of the questionnaire, the main threat is the insertion of ambiguous and leading questions. We made our best to avoid this problem by constantly reviewing and improving the proposed questions. Additionally, we ran a pilot survey to identify and correct this type of questions. In some questions, we also gave the participants the opportunity to respond with an answer different from the proposed ones by adding an \aspas{Other} option. Furthermore, with the exception of the background questions, no question was mandatory. Therefore, participants were not forced to provide answers when they did not want or when they were not sure about their answers.

There are also two threats related to the method used to retrieve the participants' e-mails. The first one is related with the match between the Stack Overflow profile and the GitHub profile of the participants. It is possible that a Stack Overflow user has been matched with a homonym user in GitHub (i.e., users who have exactly the same login name, but who are different people). Additionally, it is possible that the heuristic used to assess and rank the expertise of the selected developers does not reflect the reality. 
%Consequently, we may have removed experienced users from the top of the rank or we may have included novice users in the top of the rank.

There are also some aspects that limit the generalization of our results (external validity). First, the sample for the survey was selected only from the Stack Overflow community. Therefore, it is possible that the findings in this study do not apply to a different population. Moreover, constant and rapid changes in Web development environments, including new technologies and new versions of \AngularJS, can lead to different results if the study is repeated in the future.

Finally, we have to mention threats related to human behavior. For instance, we can mention the ordering of the questions since one may provide context for the next one. Another threat is the attitude of the participants towards the topic of research. Their responses can introduce bias to make \AngularJS\ appear in a positive or negative light. We can mention the case of one participant who declared that he is a contributor of the \AngularJS\ project.

\section{Related Work} \label{sec:relatedwork}

Some works have been focusing on practitioners' use of known technologies.
\citet{dobing2006} conducted the first survey on how UML diagrams are used by practitioners and their clients.
The authors gathered 182 responses from analysts with average experience of 4.7 years in UML.
Class diagrams are being used regularly by the majority of the respondents,
followed by Sequence and Use Case diagrams.
Some of the reasons why a UML diagram is not used vary from
\aspas{not well understood by analysts} to \aspas{insufficient value to justify the cost}.
In another study, \citet{petre2013} reported two years of interviews with practitioners.
Most of them (35 out of 50) currently do not use UML at all, due to notation overhead, lack of context, etc.
%Others use in a selective and informal way, for as long as it is considered useful;
%to capture the design as it stabilizes and generate code automatically, and
%to generate diagram from code specifically for management or client's request.
%
Both work highlight the complexity of UML.
They also suggest that more tooling is needed for both newcomers and professionals
in order to use the language more effectively.
%Our survey focuses on a technology that is relatively new, \textit{i.e.}, almost six years off its initial release,
%nevertheless we were able to gather an expressive number of respondents.
%Tool support is also suggested to assist AngularJS developers in practice.

%Previous work has been dedicated to identify inconsistencies particular to Javascript MVC applications.

\citet{ocariza2015detecting} proposed two types of inconsistencies that can be found in \AngularJS\ applications:
(i) when identifiers used in one layer are undefined in the lower layer; and 
(ii) when values assigned to a variable, or returned by a function, do not match their type in the view.
According to the authors, both inconsistencies are not easily caught during development and might cause bugs.
%The authors also proposed a tool, called Aurebesh, to automatically identify these inconsistencies.
%The approach was able to find 15 inconsistencies in 11 applications, all of them corresponding to real-world bugs.
However, in our survey, only 39\% of the respondents considered that \textit{silent failures},
corresponding to identifier inconsistency, are real problems in their daily development.
Moreover, 84.5\% of the respondents considered two-way data binding,
which relates to type consistency, as a valuable feature.
Both results shed light over real problems developers face when they \mbox{use \AngularJS}.

\section{Conclusion}
\label{chap:conclusion}

This paper reported an empirical study about different aspects of \AngularJS\
based on opinions and experiences of  developers. Our main contributions include the identification of the most appreciated features of \AngularJS\ (e.g.,~custom interface components, dependency injection, and two-way data binding) and the most problematic aspects of the framework (e.g.,~performance degradation and implementation of directives).
Future work includes an analysis of the results using statistical tests. Interviews with \AngularJS\ developers can also contribute to strengthen our findings.

\section*{Acknowledgments}
Our research is supported by FAPEMIG and CNPq. We also deeply thank the 460 developers who answered the survey.

\bibliographystyle{abbrvnat}

%\begin{spacing}{0.8}
\bibliography{bibfile}
%\end{spacing}

\end{document}